\documentclass[aps,twocolumn,prl,preprintnumbers,amsmath,amssymb,superscriptaddress]{revtex4-1}

\usepackage{graphicx}
\usepackage{bm}
\usepackage{hyperref}
\usepackage{xcolor}
\usepackage{IEEEtrantools}
\usepackage{float}

\restylefloat{table}

\begin{document}

\title{Terahertz-driven Local Dipolar Correlation in a Quantum Paraelectric}

\author{Bing Cheng}\email {chengbing986@gmail.com}
\affiliation{Stanford Institute for Materials and Energy Sciences, SLAC National Accelerator Laboratory, Menlo Park, CA 94025, USA}

\author{Patrick L. Kramer}
\affiliation{Laser Science and Technology, SLAC Linear Accelerator Laboratory, Menlo Park, CA 94025, USA}

\author{Zhi-Xun Shen}
\affiliation{Stanford Institute for Materials and Energy Sciences, SLAC National Accelerator Laboratory, Menlo Park, CA 94025, USA}
\affiliation{Geballe Laboratory for Advanced Materials, Stanford University, Stanford, CA 94305, USA}
\affiliation{Departments of Physics and Applied Physics Stanford University, Stanford, California 94305, USA}

\author{Matthias. C. Hoffmann}\email{hoffmann@slac.stanford.edu}
\affiliation{Laser Science and Technology, SLAC Linear Accelerator Laboratory, Menlo Park, CA 94025, USA}

\date{\today}

\begin{abstract}

Light-induced ferroelectricity in quantum paraelectrics is a new avenue  of achieving dynamic stabilization of hidden orders in quantum materials. In this work, we explore the possibility of driving transient ferroelectric phase in the quantum paraelectric KTaO$_3$ via intense THz excitation of the soft mode. We observe a long-lived relaxation in the THz-driven second harmonic generation signal (SHG) that lasts up to 20 ps at 10 K which may be attributed to light-induced ferroelectricity. Through analyzing the THz-induced coherent soft-mode oscillation and finding its hardening with fluence well described by a single well potential, we demonstrate intense THz pulses up to 500 kV/cm cannot drive a global ferroelectric phase in KTaO$_3$. Instead, we find the unusual long-lived relaxation of SHG comes from a THz-driven moderate dipolar correlation between the defect-induced local polar structures. We discuss the impact of our findings on current investigations of the THz-induced ferroelectric phase in quantum paraelectrics.

\end{abstract}

\maketitle

In quantum paraelectrics (QPEs), e.g. SrTiO$_3$, complete softening of their soft polar modes is prevented by the nuclear zero-point vibration\cite{QFE_1979}. Such quantum fluctuation hinders QPEs to develop long-range ferroelectric orders down to millikelvins even if a shallow ferroelectric (FE) double-well potential indeed forms below the FE Curie temperatures\cite{QFE_theory1,QFE_1979}. In decades there has been strong interest in recovering the hidden FE phase in QPEs by means of doping and straining\cite{STO_doping_1984,STO_O18_1999,STO_strain_2004}. Until recently, rapid advances in ultrafast laser technology have opened the possibility of driving QPEs into transient FE phase by intense laser pulses\cite{light_FE_theory_2017}. However, the experimental evidence is ambiguous and precise mechanism remains debated.  According to a quantum-mechanical model developed for photoexcited phases in SrTiO$_3$, the resonant THz excitation of FE soft mode could tune the lattice strain in a non-mechanical fashion, which transiently produces deep double-well potential along the soft-mode coordinate and drives a mixture of the lattice wavefunctions of the ground and first excited states. The mixture creates a nonzero expectation value of the soft-mode displacement and triggers a transient ferroeletricity\cite{THz_FE_theory1}. This transient FE phase breaks global lattice inversion symmetry and contributes an ultrafast and long-lived second harmonic generation (SHG) relaxation to the time-resolved SHG measurement, which has been claimed to be observed in SrTiO$_3$ by recent experimental works\cite{Li1079,Nova1075}.

\setlength{\parskip}{0em}

Despite the exciting progress of THz-induced transient FE phase in QPEs, some experimental challenges still exist. First, although the long-lived SHG relaxation has been observed experimentally and interpreted as the fingerprint of a transient FE phase\cite{Li1079}, the temporal dynamics of FE soft mode, a more intrinsic landmark of transient FE phase, is still uncharted. In a double-well lattice potential (ferroelectric phase), the ladder of vibrational energy eigenstates shows decreasing energy spacings as the eigenstate's index increases\cite{STO_SF_nonlinear_2021}. Intense THz excitation of soft mode will populate excited lattice states and yield a redshift of the soft mode frequency. On the other hand, a THz-driven FE phase is predicted to accompany with a light-induced deepening of the double-well potential\cite{THz_FE_theory1}, transiently increasing the energy spacings between the vibrational eigenstates. Therefore, as driven into a transient FE phase, it is not yet clear whether the FE soft mode in QPEs will soften, harden, or exhibit more complicated temporal behaviors. Second, defects and impurities usually cannot be avoided in crystals. In perovskite-type QPEs which are highly polarizable hosts, unavoidable defects and impurities always carry dipolar entities which can polarize nearby region and develop to polar nanoregions (PNRs)\cite{Samara_2003_nanodomain_review}. At the temperatures far above FE Curie temperature, these PNRs are small and behave like non-interacting point-like dipoles\cite{Samara_2003_nanodomain_review}. As temperature decreases, the size and amount of PNRs grow. Even if a global FE order does not develop in QPEs, these randomly oriented PNRs can be aligned by THz pulses and contribute signals to THz-pumped SHG probe measurement, complicating the observation of an intrinsic light-induced FE phase. 

In this work, we used THz pump SHG probe to study the quantum paraelectric KTaO$_3$ and revisited the problem of THz-induced FE phase in QPEs. KTaO$_3$ is a counterpart of SrTiO$_3$, but has a very low FE Curie temperature $T_c$ $\sim$ 4 K extrapolated from the Curie-Weiss fit to the high-temperature permittivity\cite{KTO_Tc_2016}. Unlike SrTiO$_3$, KTaO$_3$ does not experience any structural phase transitions below room temperature down to millikelvins (Fig. 1(a))\cite{KTO_cubic_1989}, avoiding the complicated phonon dynamics in SrTiO$_3$\cite{STO_105K_1969,softmode_review_1974,STO_105K_2000}. Through analyzing the THz-induced SHG signal, we demonstrated THz field up to 500 kV/cm cannot drive a global transient FE phase in KTaO$_3$. Instead, we found intense THz field drives moderate dipolar correlation between local polar nanoregions, contributing remarkable long-lived SHG relaxation persisting up to 20 ps to the time-resolved SHG measurement.

\setlength{\belowcaptionskip}{-0.5cm} 

\begin{figure}[t]
\includegraphics[clip,width=3.4in]{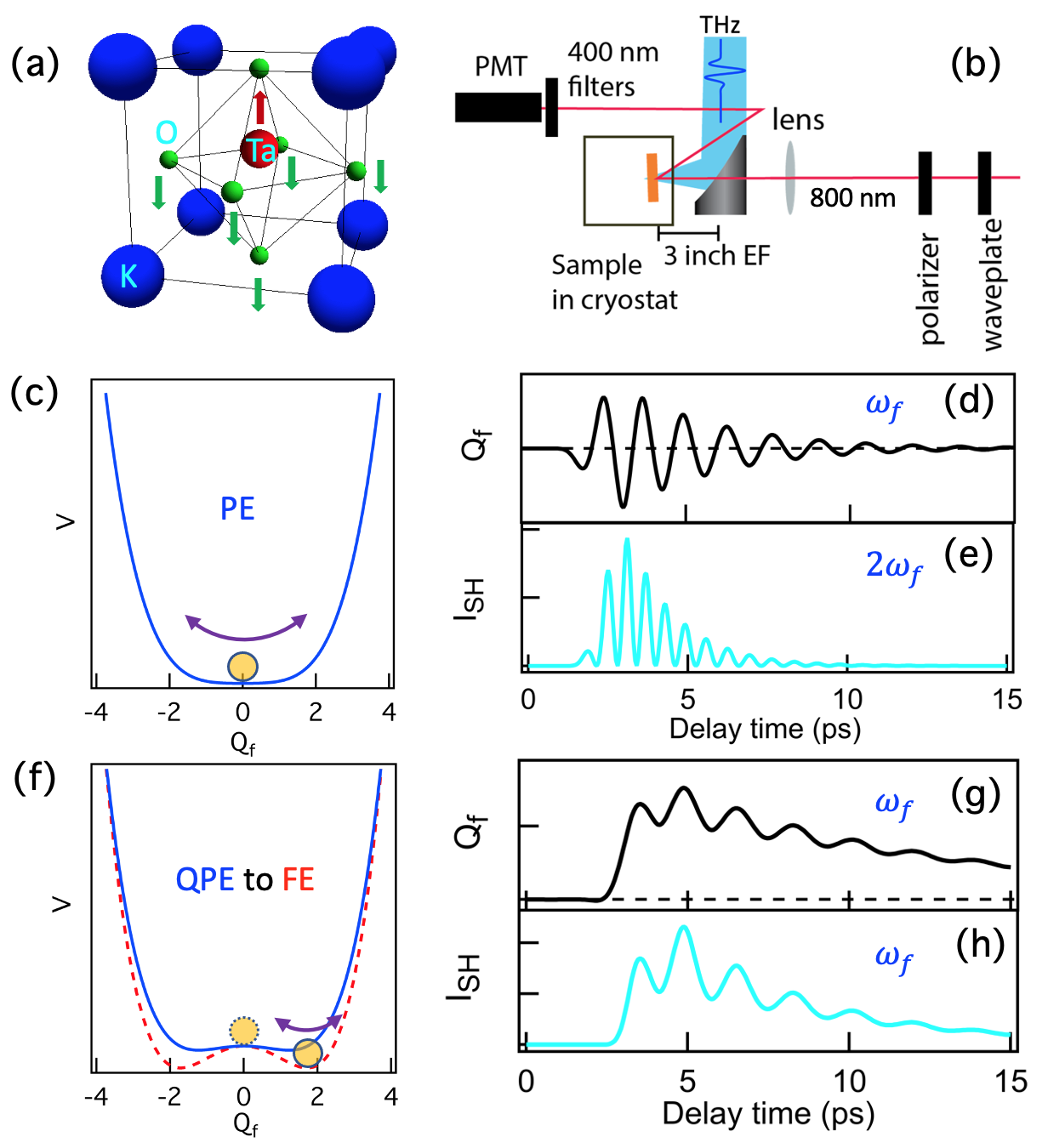}
\caption{(a) The cubic lattice structure and ferroelectric soft mode of KTaO$_3$. (b) Schematic illustration of our THz pump SHG probe setup. (c) Illustration of the single-well lattice potential along the soft-mode coordinate Q$_f$ in paraelectric (PE) state. (d) and (e) Illustration of THz-driven soft-mode displacement Q$_f$ and THz-induced SHG intensity $I_{\rm{SH}}$ as a function of delay time in the lattice potential shown in (c). Note that $I_{\rm{SH}}$ $\propto$ $Q^2_f(t)$. Hence, the oscillating frequency of $I_{\rm{SH}}$ is doubled of the fundamental soft-mode frequency. (f) Illustration of the ultrafast deepening of the FE potential along the soft-mode coordinate Q$_f$ in a THz-driven transient FE phase. Note that the soft mode will be driven to oscillate in a new potential minimum. (g) and (h) Illustration of possible soft-mode displacement Q$_f$ and THz-induced SHG signal as a function of delay time in a THz-driven transient FE phase. Because of driven to a new potential minimum, Q$_f$ will include a non-oscillatory component. The oscillation of $I_{\rm{SH}}$ follows the oscillation of Q$_f$ and will not show a simple frequency doubling.}
\label{Fig1}

\end{figure}

\setlength{\belowcaptionskip}{-0.5cm} 
 The soft mode in KTaO$_3$ is an infrared-active phonon mode (Figs. \ref{Fig1}(a)) and downshifts to $\sim$0.8 THz below 50 K\cite{KTO_softmode_2,KTO_STO_softmode_1968,Softmode_frequency_2005}. Thus, THz pulses can resonantly drive large coherent soft-mode oscillation which transiently breaks global lattice inversion symmetry and introduces a time-dependent second order susceptibility $\chi^{(2)}$\cite{chi_2_2017}. We used single-cycle THz pulses up to 500 kV/cm generated from optical rectification in LiNbO$_3$ to excite a KTaO$_3$ single crystal. The resultant SHG at 400 nm was probed by a 130 fs, 800 nm pulse as a function of time delay with respect to the THz pump pulse. The schematic of our THz pump SHG probe setup is shown in Figs. \ref{Fig1}(b). More details could be found in the supplementary material (SM).

\setlength{\belowcaptionskip}{-0.5cm}

\begin{figure}[t]
\includegraphics[clip,width=3.6in]{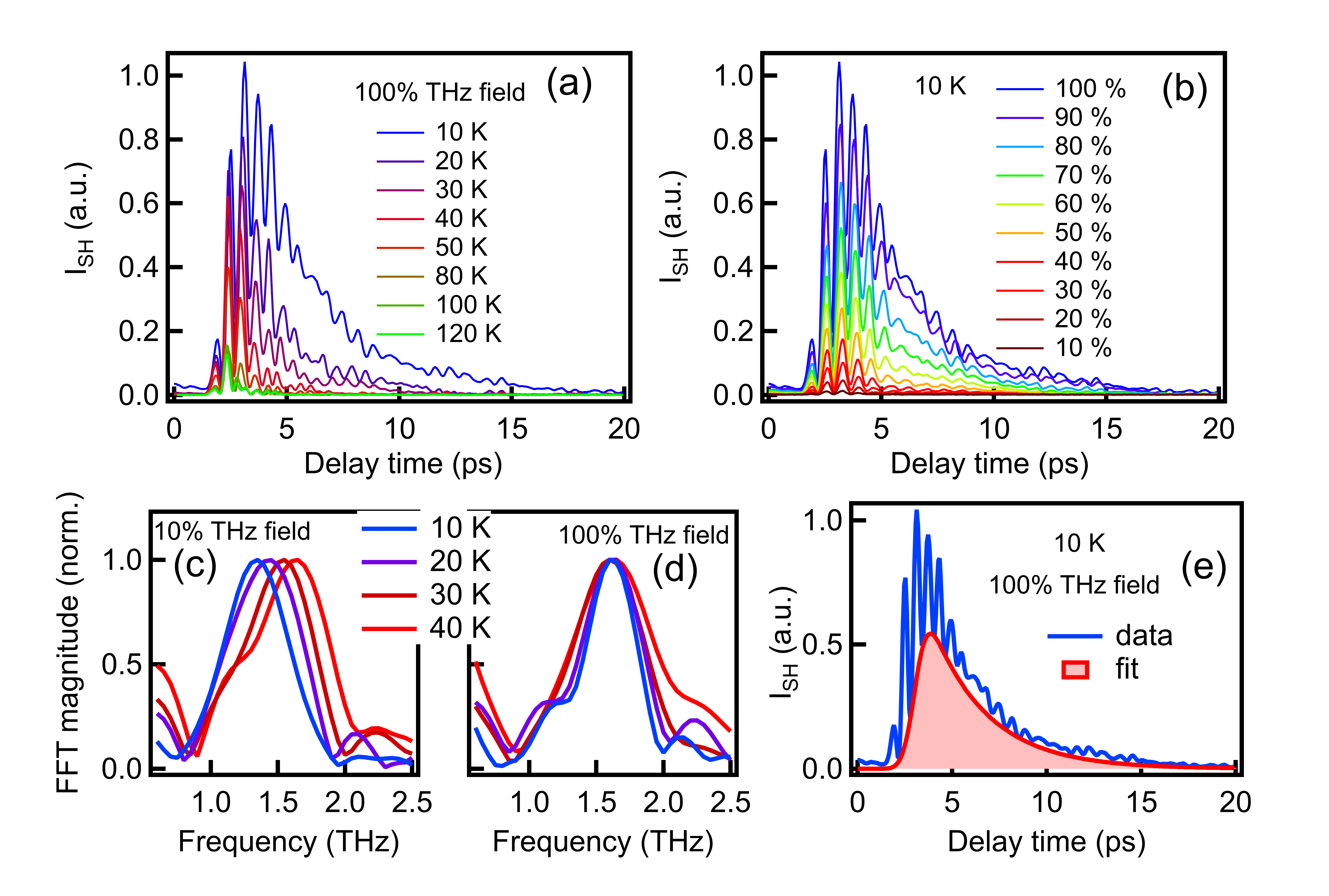}
\caption{(a) Temperature dependent time-resolved SHG signals recorded at maximum THz field strength. (b) THz field strength dependent time-resolved SHG signals at 10 K. The THz field is attenuated in 10\% steps. (c) and (d) Temperature dependent Fourier transform of time-resolved SHG signal pumped with 10\% and 100\% of maximum THz field strength respectively. (e) A single exponential relaxation fit to the non-oscillatory SHG relaxation.}
\label{Fig2}
\end{figure}

\setlength{\belowcaptionskip}{-0.5cm} 

Figs. \ref{Fig2}(a) shows THz-induced SHG intensity as a function of delay time at various temperatures. At temperatures above 80 K, the transient SHG signal is weak and primarily follows the waveform of THz pulse \textit{squared}. As temperature decreases, the SHG signal grows rapidly. In the meanwhile, a clear oscillatory feature develops on top of a non-oscillatory SHG background. At 10 K, THz field drives large SHG oscillations, as well as a significant non-oscillatory SHG component which persists to 20 ps. Similar long-lived non-oscillatory SHG relaxation has been observed in SrTiO$_3$ and interpreted to the signature of a transient FE order\cite{Li1079}. We will discuss the origin of the long-lived SHG relaxation in KTaO$_3$  later. Figs. \ref{Fig2}(c) shows a fast Fourier transform analysis of SHG oscillations at 10\% of full THz field strength. At 40 K, a single mode at 1.6 THz emerges. As temperature decreases, this mode continuously softens, consistent with the well known behavior of the KTaO$_3$ soft mode \cite{KTO_softmode_2,KTO_STO_softmode_1968,Softmode_frequency_2005}.  Figs. \ref{Fig2}(d) shows temperature dependent Fourier transforms of SHG oscillations under a full THz excitation. The mode frequencies blueshift comparing to the data in Figs. \ref{Fig2}(c). Note that in our measurements, the frequency of SHG signal is twice of the fundamental soft-mode frequency\cite{KTO_softmode_2,KTO_STO_softmode_1968,Softmode_frequency_2005}. This frequency doubling is expected in a centro-symmetric system with single well potential. As shown in Fig. \ref{Fig1}(c), if an oscillator is driven in a single-well lattice potential, in each cycle, the driven displacement $Q_f(t)$ will cross $Q_f$ = 0 twice (Figs. \ref{Fig1}(d)). Because of the SHG intensity $I_{\rm{SH}}$(\textit{t}) $\propto$ $\lvert$$\chi^{(2)}$(\textit{t})$\rvert^2$ $\propto$ $Q^2_f(t)$\cite{chi_2_2017}, the frequency of SHG oscillation will be doubled of the oscillating frequency of driven displacement Q$_f(t)$(Fig. \ref{Fig1}(e)). 

\begin{figure}[t]
\includegraphics[clip,width=3.3in]{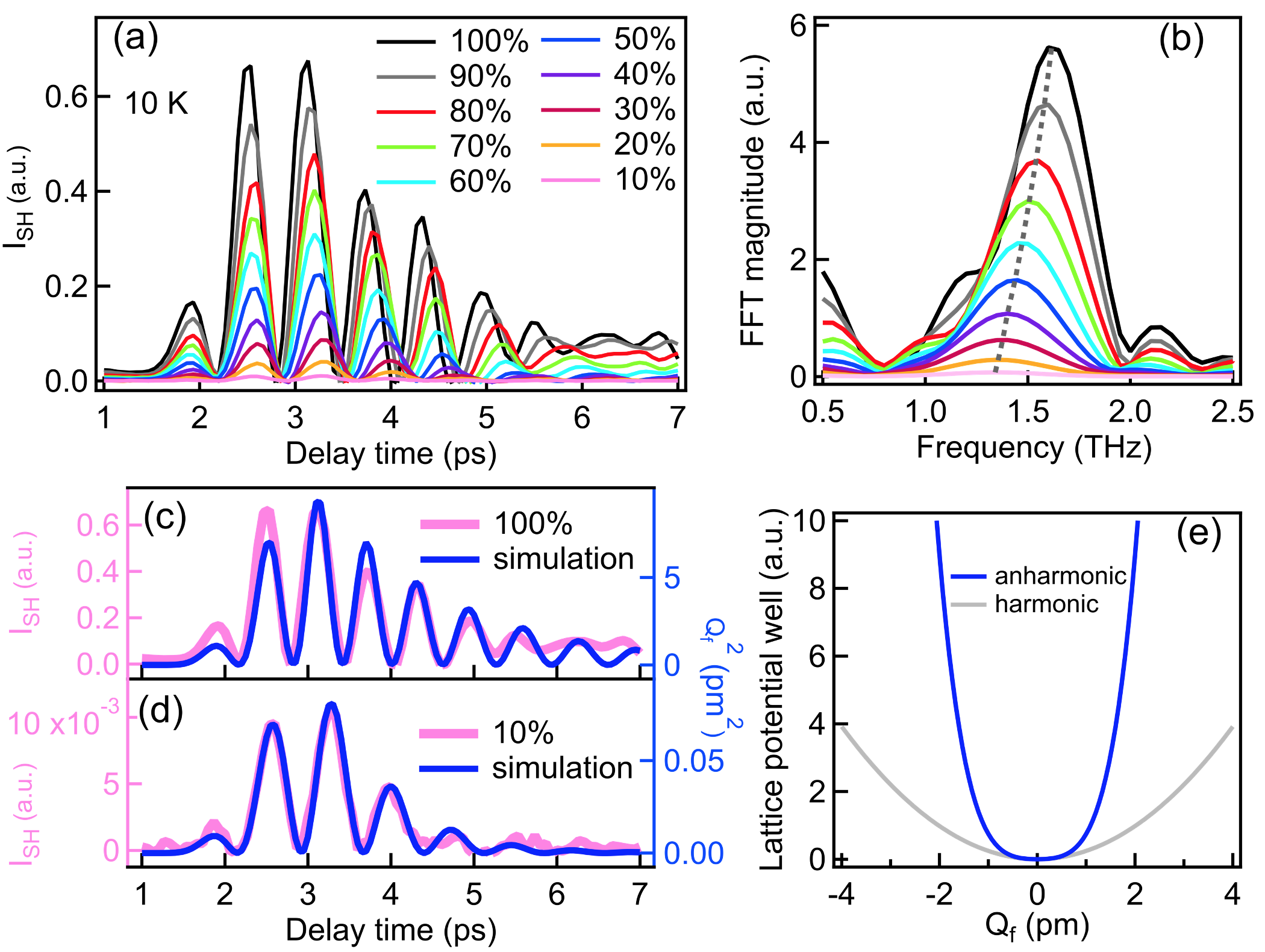}
\caption{(a) The pure oscillatory SHG signals recorded under various THz field strengths at 10 K. The maximum (100\%) THz field strength is 500 kV/cm. (b) The Fourier transforms of oscillatory SHG signals in (a). The gray dashed curve indicates the hardening trend of FE soft mode as increasing THz field strength. (c) and (d) Numerical simulations of THz-driven soft mode displacement \textit{squared} Q$_f^2$ using Eq. \ref{anharmonic} pumped by 100\% and 10\% of maximum THz field strength respectively. Due to $I_{\rm{SH}}$ $\propto$ $Q^2_f(t)$, $I_{\rm{SH}}(t)$ (left axis) and $Q^2_f(t)$ (right axis) have similar temporal waveform. (e) The experimentally reconstructed anharmonic potential of the FE soft mode plotted along with the harmonic potential. }
\label{Fig3}
\end{figure}

In Figs. \ref{Fig2}(b), we show THz-induced SHG signal at 10 K under various THz field strengths (in 10\% step). One can see the oscillatory and non-oscillatory components of SHG signal both rise rapidly with increasing pump field strength. The SHG signal is tied to temporal evolution of symmetry changes and carries intrinsic information of the soft-mode lattice potential. To isolate the pure oscillatory component, we used a single exponential relaxation function exp$(-t/\tau_0)$ convolving with the step function to capture the non-oscillatory SHG background. One simulation example was displayed in Figs. \ref{Fig2}(e). Note that the oscillatory component of SHG signal is proportional to $Q^2_f(t)$ and always positive. Hence, the simulating curve for the non-oscillatory background should capture the bottoms of the oscillation on SHG signal. After subtracting the non-oscillatory component, we plot the pure oscillatory SHG components and their Fourier transforms at 10 K in Figs. \ref{Fig3}(a) and \ref{Fig3}(b) respectively. One remarkable feature one can quickly notice in both plots is the hardening of the soft mode as the THz field strength increases, which is a typical characteristic of a driven oscillator in an anharmonic lattice potential\cite{KTO_potential_1995}. In a quartic anharmonic lattice potential $V(Q_f)=\frac{1}{2}M\omega^2_{f0}Q^2_f+\frac{1}{4}MkQ^4_f$ (Figs. \ref{Fig1}(c)), the soft-mode frequency approximately follows $\omega_f^2$ = $\omega^2_{f0}$+3$k$$Q^2_f$\cite{KTO_potential_1995}. Here $\omega_{f0}$ is the soft mode frequency in the harmonic limit. $M$ is the soft-mode reduced mass. $k$ is the coefficient of the quartic term, and $Q_f$ is the field driven soft-mode displacement. As $Q_f$ increases, the soft mode will naturally harden. In our measurement we cannot measure $Q_f$ directly, but the THz-driven $Q^2_f(t)$ is proportional to the oscillatory SHG signal $I_{\rm{SH}}(t)$. Therefore, $I_{\rm{SH}}(t)$ and $Q^2_f(t)$ should have similar temporal waveform in time domain. We numerically solved the soft-mode motion equation:

 \vspace{-0.5cm}

\begin{equation}
\frac{d^2Q_f}{dt^2}+\gamma\frac{dQ_f}{dt}+\omega^2_{f0}Q_f+kQ^3_f=\frac{Z^*eE_{in}}{M}
\label{anharmonic}
\end{equation}

\noindent Here $\gamma$ is the damping coefficient. $Z^*$ = 8.6 is the soft-mode Born effective charge\cite{Born_effective_charge_2010}. $e$ is the charge of electron. $E_{in}$ is the THz field inside KTaO$_3$. The reduced mass $M$ is determined by $[1/(M_{\rm{K}}+M_{\rm{Ta}})+1/(3M_{\rm{O}})]^{-1}$, and $M_{\rm{K}}$, $M_{\rm{Ta}}$, and $M_{\rm{O}}$ are the atomic masses of potassium, tantalum and oxygen\cite{STO_SF_nonlinear_2021}. More simulation details could be found in SM. We plot oscillatory SHG data and the numerically simulated $Q^2_f$($t$) together in Figs. \ref{Fig3}(c) and \ref{Fig3}(d). One can see the simulations well reproduce the waveforms of oscillatory SHG signal in time domain. The harmonic soft-mode frequency $\omega_{f0}$/2$\pi$ and the quartic anharmonic coefficient $k$ were found to be 0.7 THz and 2.0 pm$^{-2} $THz$^2$ respectively. The damping coefficient $\gamma$ decreases as THz field increases and was found to be $\sim$ 2$\pi$$\times$0.11 THz under a full THz excitation. The experimentally reconstructed soft-mode anharmonic potential was plotted in Fig. \ref{Fig3}(e) along with the harmonic potential of FE soft mode, similar to the single-well soft-mode lattice potential reconstructed in a 300-nm-thick SrTiO$_3$ film utilizing intense THz transmission spectroscopy\cite{THzpump_STO}.

Our analysis of the coherent soft-mode oscillation has provided critical insights on the origin of the observed long-lived and non-oscillatory SHG relaxation in KTaO$_3$. We had showed that a single-well quartic lattice potential model (Eq. \ref{anharmonic}) can well reproduce all 10 K SHG oscillations. Most importantly, the quartic coefficient $k$ of the potential is found not sensitive to the THz field strength (see SM). The potential profile does not experience detectable changes as pump field strength increases to 500 kV/cm. Moreover, in such an anharmonic potential (Fig. \ref{Fig3}(e)), the THz-driven soft-mode hardening is actually adverse to stabilize a FE phase. The large THz-driven magnitude of $Q^2_f(t)$, which acts like an enhanced thermal-averaged $\langle$Q$^2_f$$\rangle$ by elevating temperature\cite{KTO_potential_1995}, should drive KTaO$_3$ into a deep paraelectric phase. In this regard, the non-oscillatory component of SHG signal, despite persisting to 20 ps at 10 K, is unlikely from a global light-induced FE phase transition. 

\begin{figure}[t]
\includegraphics[clip,width=3.3in]{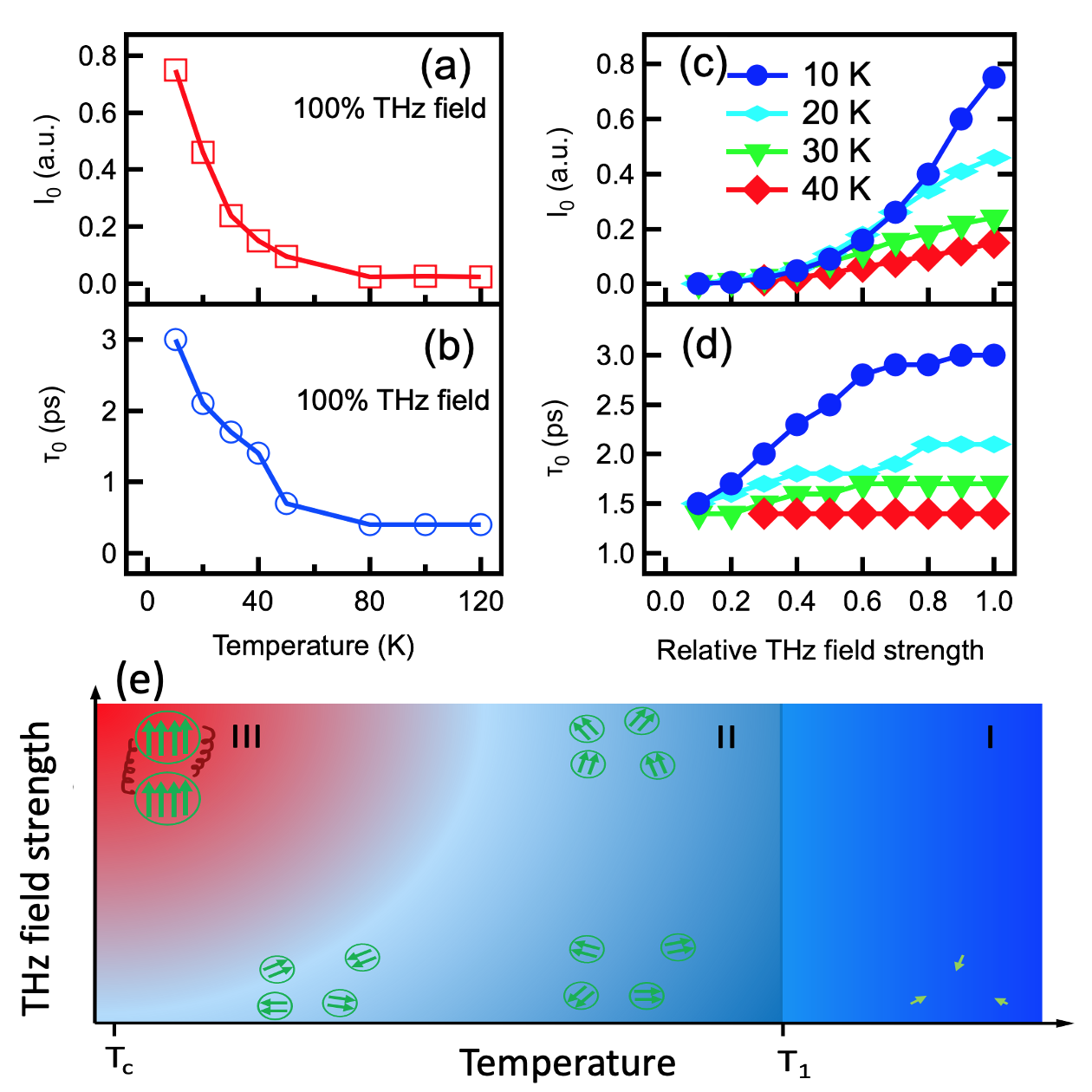}
\caption{(a) and (b) The pump-probe magnitude $I_0$ and relaxation time $\tau_0$ of THz-induced non-oscillatory SHG component as a function of temperature under maximum THz excitation. (c) and (d) The pump-probe magnitude $I_0$ and relaxation time $\tau_0$ of THz-induced non-oscillatory SHG component as a function of THz field strength at different temperatures. (e) A qualitative phase diagram of PNR dynamics in the THz-pumped KTaO$_3$. $T_1$ $\sim$ 60 K is the onset temperature of PNRs.  $T_c$ $\sim$ 4 K is the FE Curie temperature. Above $T_1$, as shown in Region I, the point-like dipoles are dominant. Below $T_1$, PNRs start to appear. In Region II, these PNRs behave like non-interacting dipolar entities even if they are partially polarized by strong THz pulses. As temperature is further lowered, as shown in Region III, strong THz pulses not only polarize PNRs, they also enhance the dipolar interaction between PNRs and induce longer depolarization relaxation. }
\label{Fig4}
\end{figure}

To explore its origin, we employed detailed analysis of the non-oscillatory SHG component. We used a single exponential relaxation function to extract the pump-probe amplitude $I_0$ and relaxation time $\tau_0$ (Figs. \ref{Fig2}(d)). More fitting details can be found in SM. We plot $I_0$ and $\tau_0$ as a function of temperature under maximum THz field excitation in Figs. \ref{Fig4}(a) and \ref{Fig4}(b). At temperatures above 80 K, $I_0$ is negligible and $\tau_0$ is $\sim$400 fs. As temperature decreases, both quantities surge. Their onset temperatures, $\sim$60 K, coincide with the temperature reported for local polar structures arising in KTaO$_3$\cite{KTO_dielectric_2014,Samara_2003_nanodomain_review}. 

It is well established that even if in nominally pure perovskite-type QPEs, the unavoidable impurities and defects usually bring in dipolar entities which could develop to local polar structures at the nanometer scale\cite{Samara_2003_nanodomain_review}. In pure KTaO$_3$, the defects are mainly from the Ta$^{+3}$/Ta$^{+4}$ and oxygen vacancies\cite{KTO_defect_1996,KTO_defect_1997} which introduce polar nanoregions (PNRs) below $\sim$50 K\cite{nanodomain_2015,KTO_dielectric_2014}. These PNRs break local lattice inversion symmetry but orient randomly, which has been reported to contribute a weak and incoherent SHG background signal to static SHG measurement below 50 $\sim$ 60 K\cite{KTO_defect_1996,KTO_SHG_2001}.  In our measurement, the randomly orientated PNRs can be transiently polarized by THz electric field and contribute a large initial amplitude to the THz-induced SHG signal. As THz pulses are turned off, the polarized PNRs will gradually depolarize, resulting in a temporal SHG relaxation as a function of delay time.

Figure \ref{Fig4}(c,d) shows the THz-induced SHG amplitude $I_0$ and relaxation time $\tau_0$ as functions of THz field strength.  At 40 K,  $I_0$ grows slowly and linearly with increasing THz field, and $\tau_0$ is insensitive to THz field strength. The increase of the SHG amplitude seems to not modify the relaxation process. Such behavior implies the PNRs at 40 K act like non-interacting dipolar entities.  In contrast, at 10 K, $I_0$ rises in a highly nonlinear fashion as a function of THz field strength, which is quite similar to the observation in SrTiO$_3$\cite{Li1079}. The relaxation time $\tau_0$ at lowest field strength is 1.5 ps, close to the values of 40 K. As THz field strength increases, $\tau_0$ grows rapidly, and then gradually saturates at 3 ps. The large increase of $\tau_0$ strongly indicates a marked dipolar correlation between PNRs develops under high THz field excitation. It had been noted before that the polarized dipoles with close neighbors would take longer time to depolarize than normal isolated dipoles once the polarizing field is turned off\cite{dipolar_correlation_review_1990}. In our case, as more PNRs are polarized by THz field at the lowest temperature, parts of PNR pairs will tend to be bound at a temporal potential minimum of the dipole-dipole interaction Hamiltonian, which eventually prevents them to depolarize rapidly as usual and results in a longer relaxation\cite{dipolar_correlation_review_1990}. Despite more efforts required to fully understand the process, it is clear that such THz field-enhanced correlation is the primary reason for the long-lived SHG relaxation persisting up to 20 ps. 

A qualitative phase diagram of PNR dynamics in the THz-pumped KTaO$_3$ is displayed in Figs. \ref{Fig4}(e). Besides the polarizing and depolarizing dynamics of PNRs, the most striking feature is the emergence of a THz-enhanced dipolar correlation region (Region III) at low temperatures and high THz field strengths. This phase diagram is constructed from our results of KTaO$_3$, but it could be applied to other QPEs where the response of PNRs dominates.

Finally, we would like to make remarks on the general relevance of our findings to current investigations of the THz-induced FE phase in QPEs. Our work provides a standard protocol to disentangle the non-oscillatory and oscillatory components of THz-induced SHG signal. We showed a light-induced FE phase is not the only possible origin of the observed long-lived non-oscillation SHG relaxation in QPEs. The local polar structures formed at low temperatures, along with the THz-enhanced dipolar correlation between them, can also result in such a long-lived SHG relaxation persisting up to 20 ps. It is noteworthy that the amplitude of non-oscillatory SHG relaxation in KTaO$_3$ is an increase function of THz field strength (Figs. \ref{Fig4}(c)), which is distinct from the expected field evolution of the SHG relaxation in THz-induced transient FE phase. In the transient FE phase predicted by a quantum model\cite{THz_FE_theory1}, the global FE polarization comes from the THz-driven mixture of the ground and first excited states of the lattice wavefunctions. As THz field strength reaches a threshold, more higher excited lattice states will be populated. Such an over-populated and light-mixed lattice state behaves like a normal thermally excited paraelectric state which will suppress the global FE polarization\cite{THz_FE_theory1}. The THz-induced order parameter (global FE polarization), or equivalently, the non-oscillatory SHG relaxation amplitude, should show a non-monotonic field evolution (increases first and then decreases)\cite{THz_FE_theory1}. Hence, our observation is crucial and imposes strong restrictions on interpreting the THz induced long-lived SHG relaxation to the signature of a transient FE phase.

A more convincing way to demonstrate a THz-induced FE phase is to track the temporal dynamics of FE soft mode. In a light-induced FE phase, the soft mode will be driven to oscillate at a new potential minimum ($Q_f$ $\neq$ 0) transiently (Figs. \ref{Fig1}(f)). The resultant coherent oscillation of soft-mode displacement (Figs. \ref{Fig1}(g)) will not cross $Q_f$ = 0 as long as the transient FE phase has not degenerated significantly (Figs. \ref{Fig1}(f)). Therefore, the oscillatory component of SHG, despite still proportional to $Q^2_f$ (Figs. \ref{Fig1}(h)), will closely follow the oscillation of the soft-mode displacement $Q_f(t)$ rather than only showing frequency doubling (Figs. \ref{Fig1}(g)). Careful analysis of the soft-mode trajectory in time domain, and the soft-mode hardening or softening in frequency domain, as we demonstrated in KTaO$_3$, will enable to extract information of or even reconstruct the light-induced double-well potential of the transient FE phase.

In summary, we used THz pump SHG probe to study the possibility of driving a quantum paraelectric KTaO$_3$ into transient FE phase. We started by analyzing the soft-mode oscillations revealed by SHG probe and found that the observed hardening of the soft mode can be described by a single well potential, distinct from what is expected from a THz-induced FE phase. We demonstrated the long-lived non-oscillatory SHG relaxation at 10 K, despite seemingly from a transient FE phase, actually results from the THz-induced moderate dipolar correlation between the defect-induced local polar structures. Our results are in agreement with current theoretical predictions and paved the road for the realization of THz-induced FE phase in QPEs unambiguously.

We would like to thank Mengkun Liu, Xinshu Zhang and Jonathan A. Sobota for helpful discussions. Use of the Linac Coherent Light Source, SLAC National Accelerator Laboratory, is supported by the US Department of Energy (DOE), Office of Science, Basic Energy Sciences, under contract no. DE-AC02-76SF00515. The work at SIMES (BC and ZXS) is supported by the U.S. Department of Energy (DOE), Office of Science, Basic Energy Sciences, Materials Sciences and Engineering Division under Contract No. DE-AC02-76SF00515; the work at LCLS (PK and M.C.H) is supported by U.S. Department of Energy (DOE), Office of Science, Basic Energy Sciences, under award no. 2018-SLAC-100499.

 \bibliography{Quadratic}

 \newpage

\setcounter{figure}{0}
\setcounter{equation}{0}
\setcounter{section}{0}
\begin{widetext}

\maketitle

\section{S\lowercase{upplementary} N\lowercase{ote} 1:  TH\lowercase{z pump second harmonic generation} }

The schematic of our THz pump second harmonic generation (SHG) probe setup is depicted in SI Fig. 1(a). The single-cycle THz pulses were generated by optical rectification of 130 fs 800 nm laser pulses with 4 mJ energy using the tilted pulse front method in LiNbO$_3$. The THz pulses with energies up to 3 $\mu$J were focused onto the sample yielding a spot size of about 1.2 $\times$ 1.2 mm$^2$ and a fluence of $\approx$ 0.2 mJ/cm$^2$. The THz field strength at the focus was measured by electro optical (EO) sampling in a 50 $\mu$m thick 110-cut GaP crystal. The maximum field strength was 500 kV/cm. The waveform of THz pulse in time domain is displayed in SI Fig. 1(b). Note that, to calibrate the THz field strength, we used a 50 $\mu$m thick 110-cut GaP crystal as the EO sampling crystal to measure the THz pulse in time domain. The GaP crystal is very thin, so there are peaks from internal reflection following the main single-cycle THz peak. Based on the thickness and the THz refractive index of GaP crystal, these internal reflection peaks are separated by nearly 1 ps from each other. These internal reflection-induced peaks are strictly measurement artifacts in the EO sampling process. The THz pulse incident on the sample consists only the main single-cycle peak at 2 $\sim$ 4 ps.  We especially point this out here to avoid more confusion after looking at the pulse profile. The Fourier transform magnitude of the THz pulse is shown in Fig. 1(c). The KTaO$_3$ crystal we studied is 100-cut. The THz polarization is set along the 100 axis. A pair of wiregrid polarizers, controlled by a computerized rotation stage, was used to continuously attenuate the THz field. The linearity of this attenuation was verified by EO sampling. Strong probe pulses were directed onto the sample collinearly with the THz beam. The fundamental laser pulse (800nm) and the second harmonic generation pulse (400 nm) are reflected by the sample and directed through a group of filters and finally only the second harmonic generation signal was directed into photomultiplier detector.

\begin{figure}[h]
\includegraphics[clip,width=5.6in]{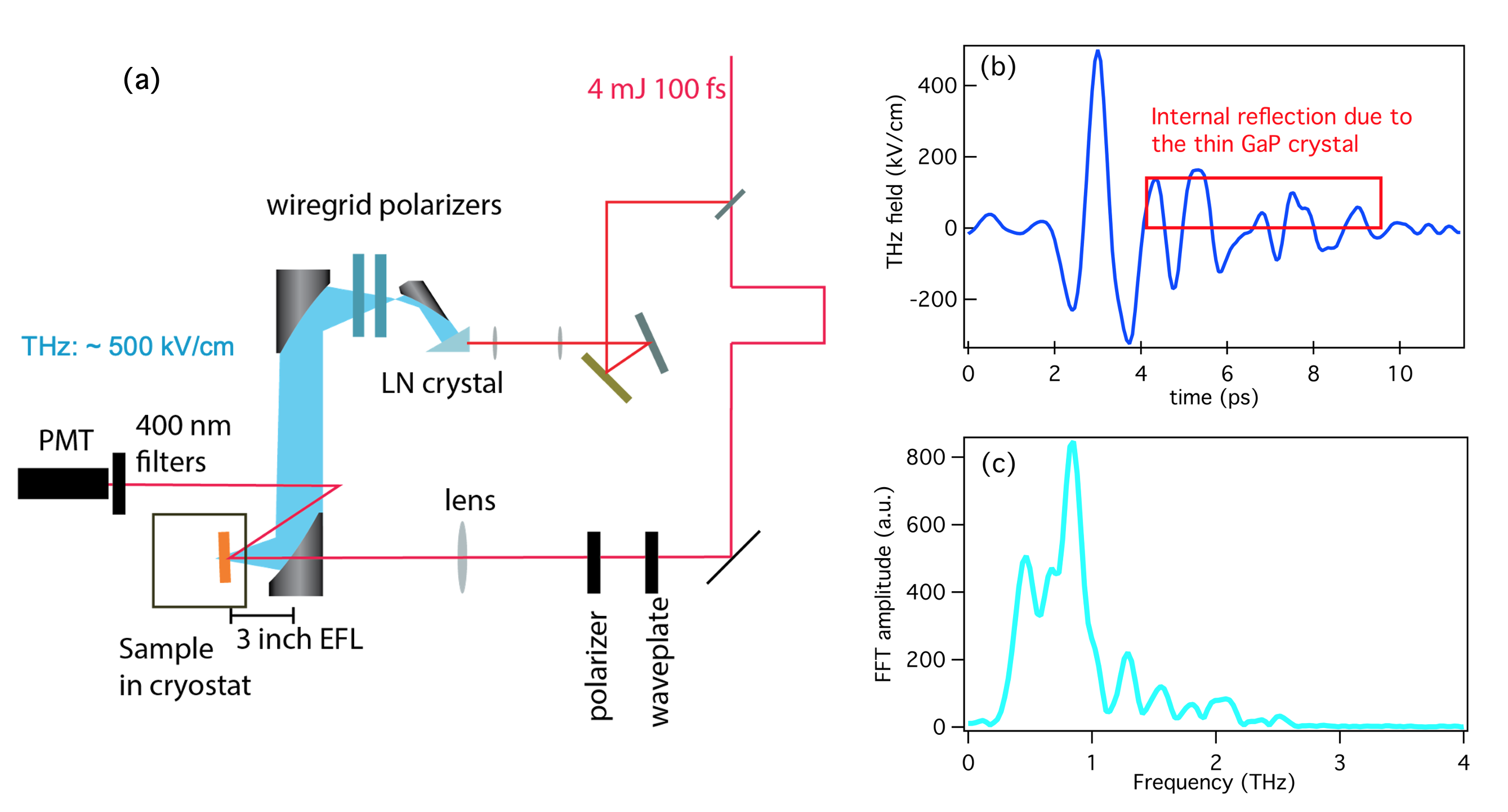}
\caption{(Color online) (a) The schematic of our THz pump second harmonic generation probe setup. (b) Electro-optic sampling signal of THz pulse measured by EO sampling in a 50 $\mu$m thick (110) GaP crystal. (c) The Fourier transform magnitude of the THz pulse.  }
\label{xxx}
\end{figure}

\begin{figure}[b]
\includegraphics[clip,width=6in]{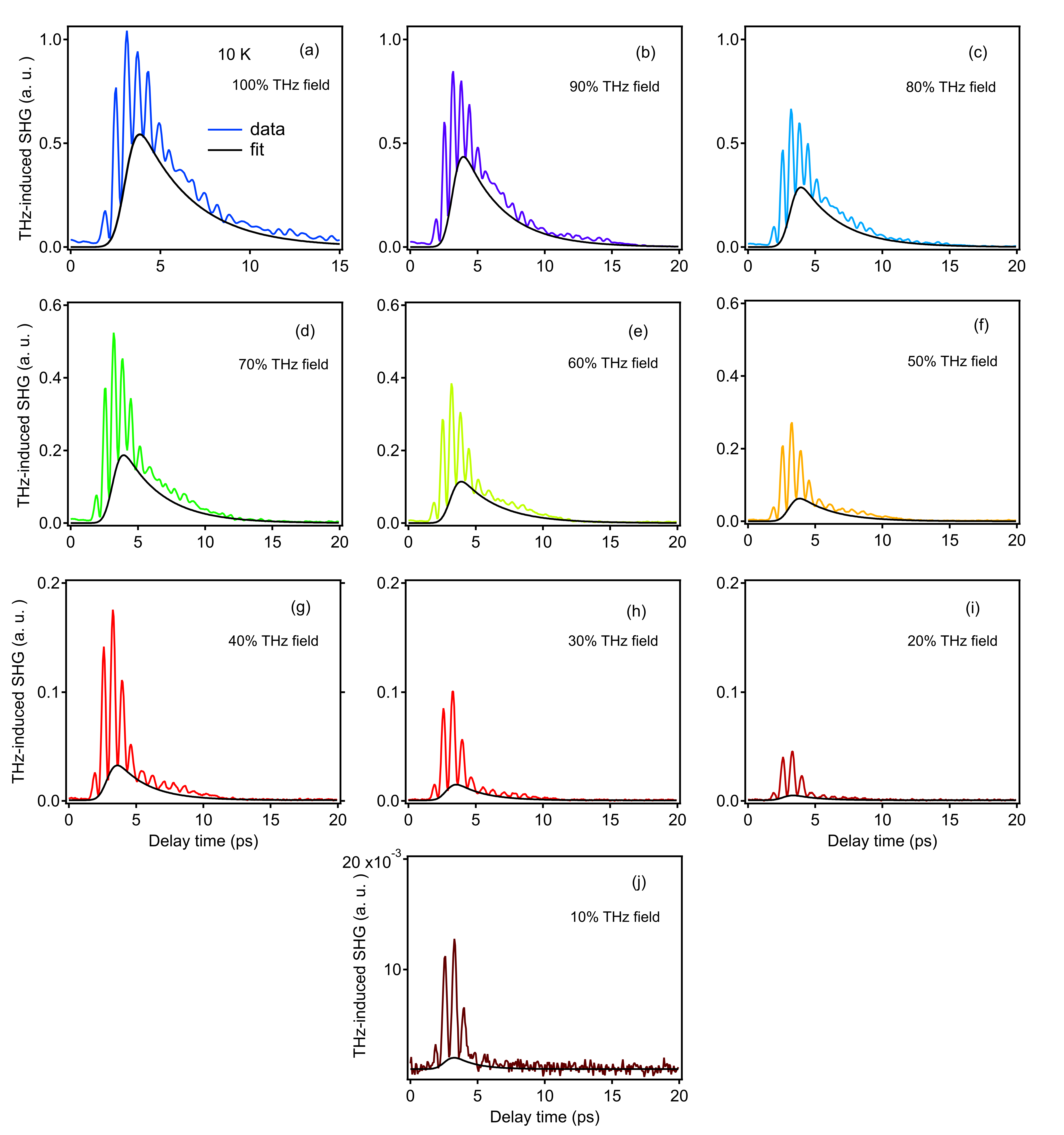}
\caption{(Color online) Fit to the non-oscillatory component of the THz-induced SHG signal at 10 K. }
\label{trans}
\end{figure}

\section{S\lowercase{upplementary} N\lowercase{ote} 2:  F\lowercase{it to the non-oscillatory component of the} TH\lowercase{z-induced} SHG \lowercase{signal} }

As we discussed in the main text, the oscillatory component of SHG signal in KTaO$_3$ should be proportional to soft mode displacement squared $Q^2_f(t)$ and always positive. Hence, the simulating curve for the non-oscillatory component should capture the bottoms of the oscillation on SHG signal. We used a simple exponential relaxation function convolving with a step function to capture the non-oscillatory SHG component.

\begin{equation}
I(t)=I_0exp(-(t-t_0)/\tau_0)(1+erf(2\sqrt{ln2}(t-t_0)/w))/2
\label{anharmonic}
\end{equation}

\noindent Here $I(t)$ is the THz-induced SHG intensity. $I_0$ is the pump-probe amplitude. $t_0$ is the time zero. $\tau_0$ is pump-probe relaxation time. $erf$ is the error function to simulate the rise of $I(t)$. $w$ is the rise width. Note that in this work, we only focus on the analysis of THz pump field dependence of the pump-probe amplitude and relaxation time. The way to simulate the rise will not modify the pump field dependence. As shown in SI Figure 2, a signal exponential decay could well describe the relaxation of the non-oscillatory component of THz-induced SHG signal at 10 K.

\begin{figure}[t]
\includegraphics[clip,width=6in]{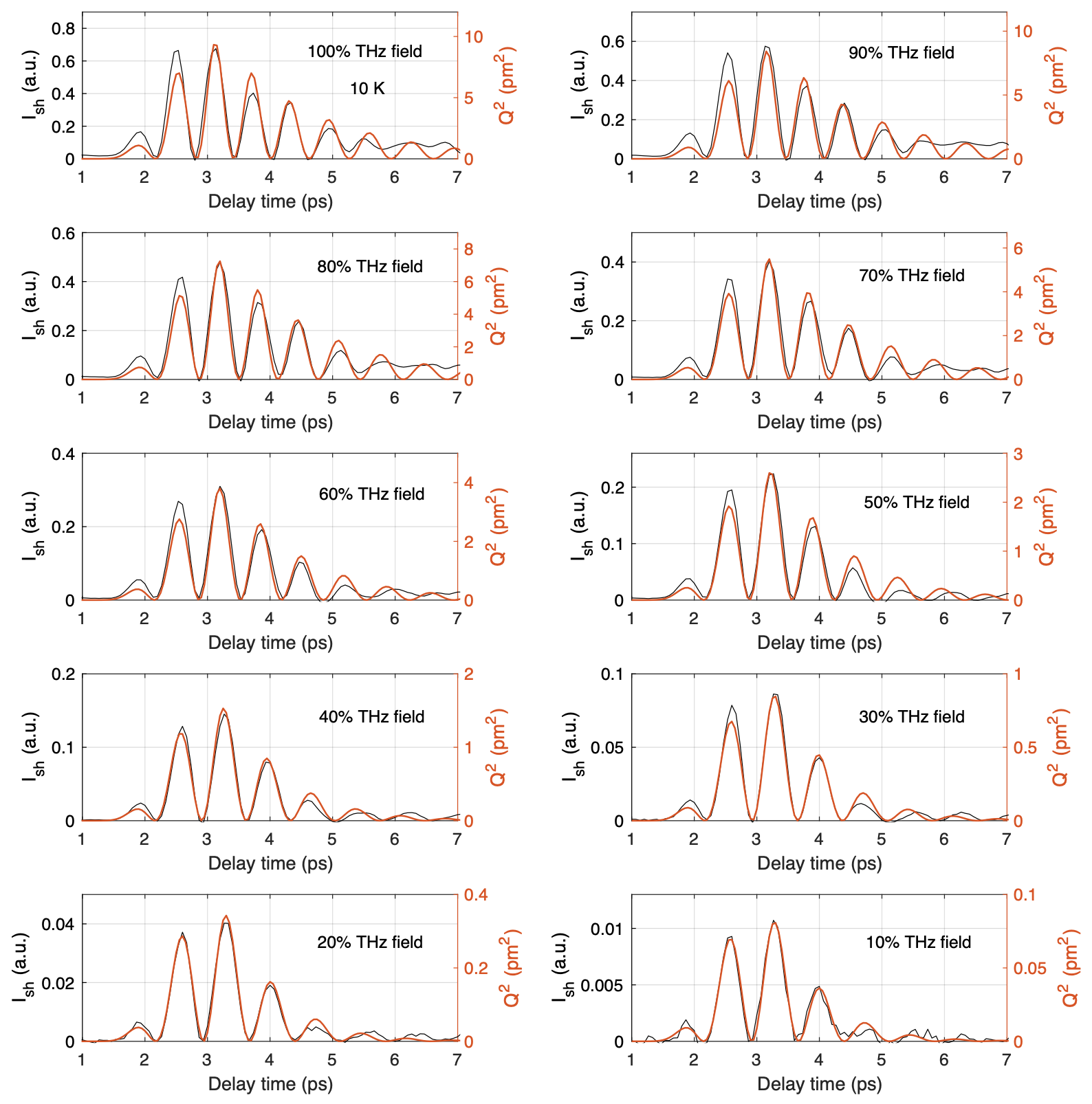}
\caption{(Color online) Simulation of the frequency hardening of soft mode in KTaO$_3$ by using a driven oscillator model.}
\label{trans}
\end{figure}

\section{S\lowercase{upplementary} N\lowercase{ote} 3:  S\lowercase{imulation of the coherent soft-mode oscillation} }

In main text, we had elaborated that the THz-induced soft mode hardening could be described by an driven oscillator model. To capture the feature of soft-mode hardening, we numerically solved the soft-mode motion equation as shown in main text. For the convenience of the simulation, we used an artificial THz pulse which could be simply described by an analytical formula as an initial input to numerically solved the Eq. 1 in main text. This artificial THz pulse mimics the real single-cycle THz pulse generated by LiNbO$_3$ crystal for our experiment as shown in SI Fig. 1(b) but rejecting the EO sampling artifacts highlighted by the red box. As we discussed in main text, our maximum field strength of the THz pulse is $\sim$ 500 kV/cm. Note that this maximum field strength is measured in the air. Inside the KTaO$_3$ crystal, the THz field strength will be decayed by the dielectric property of KTaO$_3$. Only the THz field inside the crystal could interact with lattice and drive the soft mode. To calculate the THz field strength inside KTaO$_3$, we simply used the formula $E_{in}$ =2$E_{air}$/(1+$n$). Here $E_{in}$ is the THz field inside crystal. $E_{air}$ is the THz field measured by EO sampling in the air. $n$ is the refractive index of KTaO$_3$ in THz region and is found to be $\sim$ 63. We show our simulation of the coherent soft-mode oscillation in SI Figure 3. One can see our simulations capture the main features of the soft oscillation. The simulation parameters, the soft mode frequency in the harmonic limit $\omega_{f0}$/2$\pi$, the quartic anharmonic coefficient $k$, are found to be round 0.7 THz and 2.0 pm$^{-2} $THz$^2$ respectively. The damping coefficient $\gamma$ varies from 2$\pi$$\times$0.11 THz at 100$\%$ THz field to 2$\pi$$\times$0.27 THz at 10$\%$ THz field.

\end{widetext}

\end{document}